\documentclass[12pt]{article}
\usepackage{amssymb}
\usepackage{amsmath}
\usepackage{graphicx}
\usepackage{indentfirst}
\usepackage{cite}

\begin{document}

\title{Recall a prediction on $J/\psi$ nuclear modification factor
for STAR measurements}
\author{Xiao-Ming Xu}
\date{}
\maketitle \vspace{-1cm} 
\centerline{Department of Physics, Shanghai University, Baoshan, 
Shanghai 200444, China}

\begin{abstract}
STAR collaboration has offered eminent nuclear modification factor of $J/\psi$
at high $p_T$ and midrapidity produced in Cu-Cu collisions at $\sqrt {s_{NN}}=
200$ GeV. Recalling a prediction we can understand that the feature of 
high-$p_T$ nuclear modification factor is related to $c\bar c$ produced by 
$2 \to 1$ and $2 \to 2$ partonic processes in deconfined matter particularly
in the prethermal stage and to the recombination of $c$ and $\bar c$. The
nuclear modification factor at high $p_T$ is sensitive to the earliest form of
deconfined matter that does not have a temperature.
\end{abstract}


\newpage
\vspace{0.5cm}
Recently, STAR collaboration has measured the midrapidity ratio $R_{AA}$
of $J/\psi$ produced in Cu-Cu collisions to $p$-$p$ collisions at
$\sqrt {s_{NN}}= 200$ GeV \cite{ZT}. The ratio is a 
function of the transverse momentum $p_T$. The ratio increases with increasing
$p_T$ and arrives at 0.9$\pm$0.2 at $p_T>5$ GeV/$c$. Error bars at
$p_T>5$ GeV/$c$ are large. 
If $J/\psi$, $\chi_c$ and $\psi'$ undergo only dissociation processes due to
the interaction with gluons of quark-gluon plasma \cite{KS}, the ratio must be
smaller than 1 \cite{XKSW}. On the other hand, 
the transverse momentum larger than 5 GeV/$c$ is very much higher
than the average momentum of quarks and gluons of quark-gluon plasma in thermal
equilibrium. If the $J/\psi$ nuclear modification factor
$R_{AA}$ at $p_T>5$ GeV/$c$ is taken to
be larger than 1, how can we understand the measured $p_T$ dependence? 
This is can be understood from my prediction \cite{Xu02} as follows. 

The prediction in Ref. \cite{Xu02} is about the
ratio of momentum distribution of $J/\psi$ produced in central Au-Au collisions
at $\sqrt {s_{NN}}=200$ GeV to nucleon-nucleon collisions. The predicted ratio
is shown by the solid curve in Fig. 1. The $J/\psi$ production
includes the contributions of direct $J/\psi$, the radiative feeddown from 
direct $\chi_{cJ}$ and the decay of direct $\psi'$. The theoretical ratio is
larger than 1 at the transverse momentum between 2.5 and 7 GeV at rapidity 
$y=0$. This enhancement as stated below is caused by 
$c\bar c$ yielded through $2 \to 1$ and $2 \to 2$ partonic processes in
deconfined matter particularly in the prethermal stage \cite{Xu02} and by
the recombination of charm quark and charm antiquark  
\cite{Xu99,BMS,TSR,GR,GKSG,LM,GKR,YZX}.
  
The history of a Au-Au nuclear collision at RHIC energies can be divided into 
four stages: (a) the initial nucleus-nucleus collision where quark-gluon matter
is produced; (b) the prethermal stage where quark-gluon matter thermalizes and
a temperature is eventually established; (c) the thermal stage where 
quark-gluon plasma evolves and is defined as quark-gluon matter with a
temperature; (d) evolution of hadronic matter until kinetic freeze-out. 
Longitudinal 
expansion of deconfined matter and hadronic matter is assumed in Ref. 
\cite{Xu02}. I stress that quark-gluon matter in the prethermal stage
does not have a temperature and is the earliest form of deconfined matter. 

A $c\bar c$ pair is produced in the initial nuclear collision, in the 
prethermal stage and in the thermal stage. The produced $c\bar c$ pair is a 
pointlike color singlet or a color octet pair from $2 \to 1$ processes
$a+b \to c\bar c$ and $2\to 2$ processes $a+b \to c\bar {c} +{\rm x}$ where
$a$, $b$ and $\rm x$ denote partons. The pointlike $c\bar c$ pair expands and 
may get into free states when it travels through the deconfined matter. 

A charm quark and a charm antiquark can recombine into a bound state with a 
probability.
The probability is proportional to the nonperturbative matrix elements in 
nonrelativistic QCD \cite{NRQCD},
\begin{displaymath}
<{\cal O}^H_8( ^3S_1)>,~~~~~<{\cal O}^H_8( ^1S_0)>,~~~~~<{\cal O}^H_8( ^3P_0)>
\end{displaymath}
where ${\cal O}^H_8( ^3S_1)=\chi^+ \vec {\sigma}T^a \psi \cdot (a^+_Ha_H)
\psi^+ \vec {\sigma}T^a \chi$,
${\cal O}^H_8( ^1S_0)=\chi^+ T^a \psi (a^+_Ha_H)\psi^+ T^a \chi$, and
${\cal O}^H_8( ^3P_0)=\frac {1}{3} 
\chi^+ (- \frac {i}{2}\stackrel {\leftrightarrow}{D} \cdot \vec {\sigma})T^a 
\psi (a^+_Ha_H) \psi^+ (- \frac {i}{2}\stackrel {\leftrightarrow}{D}  \cdot 
\vec {\sigma})T^a \chi$ with $\psi$ as the Pauli spinor field that annihilates 
a heavy quark, $\chi$ as the Pauli spinor field that creates a heavy 
antiquark and $a^+_H$ as the operator that creates the quarkonium $H$ in the 
out state. The nonperturbative matrix elements are constants. In the
recombination mechanism proposed in Ref. \cite{Xu99}, the probability for
$c\bar c$ to form a bound state is a constant.

When $c\bar c$ penetrates through deconfined matter, it is broken into a free
charm quark and a free charm antiquark by reactions
\begin{displaymath}
g+c\bar {c}[n ^{2S+1}L_J^{(1)}] \to c+\bar c,~~~~~
g+c\bar {c}[ ^{2S+1}L_J^{(8)}] \to c+\bar c
\end{displaymath}
where $n^{2S+1}L_J^{(1,8)}$ is the spectroscopic notation for quantum numbers 
and for singlet or octet by the superscript. Cross sections for the reactions 
were obtained \cite{Xu02}
with a formula in Ref. \cite{KS,Peskin}. In hadronic matter
charmonia are dissociated by mesons into charmed mesons via the reactions
\begin{displaymath}
q\bar {q}+c\bar {c}[n ^{2S+1}L_J^{(1)}] \to q\bar {c}+c\bar q
\end{displaymath}
Cross sections for the reactions were calculated \cite{Xu02} with a formula of 
Ref. \cite{BS}. Based on these cross sections $c\bar c$ survival probability 
can be obtained.

Momentum distribution of direct charmonium consists of five terms
\begin{equation}
\frac {dN_{\rm direct}}{dyd^2p_T} =
\frac {dN_{\rm ini}^{2 \to 2}}{dyd^2p_T}(S_{a/A} \not= 1)
+  \frac {dN_{\rm pre}^{2 \to 1}}{dyd^2p_T}
+  \frac {dN_{\rm pre}^{2 \to 2}}{dyd^2p_T}
+  \frac {dN_{\rm the}^{2 \to 1}}{dyd^2p_T}
+  \frac {dN_{\rm the}^{2 \to 2}}{dyd^2p_T}
\end{equation}
where the five terms result from $c\bar c$ pairs produced in the initial 
nuclear collision via the $2 \to 2$ processes, in the prethermal stage via the
$2 \to 1$ and $2\to 2$ processes and in the thermal stage via the $2\to 1$ and
$2\to 2$ processes, respectively. Every term is the product of two parton 
distribution functions convoluted with the product of the short-distance
production part, recombination probability and survival probability. The
momentum distribution of prompt $J/\psi$, 
$\frac {dN_{\rm prompt}^{J/\psi}}{dyd^2p_T}$, 
includes the contributions of
direct $J/\psi$, the radiative feeddown from direct $\chi_{cJ}$ and the decay
of direct $\psi'$. Let 
$\frac {dN_0^{J/\psi}}{dyd^2p_T}=
\frac {dN_{\rm ini}^{2 \to 2}}{dyd^2p_T}(S_{a/A} = 1)$ be the momentum 
distribution of prompt $J/\psi$ while the cross sections for charmonia
dissociated by gluons and hadrons are set as zero. The nuclear modification
factor is
\begin{equation}
R_{AA} = 
\frac{dN_{\rm prompt}^{J/\psi}}{dyd^2p_T} / \frac {dN_0^{J/\psi}}{dyd^2p_T}
\end{equation}
The $R_{AA}$ has been shown in Fig. 1.

$R_{AA} <1$ if $S_{a/A} \not= 1$, $\frac {dN_{\rm pre}^{2 \to 1}}{dyd^2p_T} =
\frac {dN_{\rm pre}^{2 \to 2}}{dyd^2p_T} =
\frac {dN_{\rm the}^{2 \to 1}}{dyd^2p_T} =
\frac {dN_{\rm the}^{2 \to 2}}{dyd^2p_T} =0$,
and the charmonium dissociation cross sections are taken into account. 
Therefore, $R_{AA}>1$ corresponding to
$\frac {dN_{\rm pre}^{2 \to 1}}{dyd^2p_T} \not= 0$,
$\frac {dN_{\rm pre}^{2 \to 2}}{dyd^2p_T} \not= 0$,
$\frac {dN_{\rm the}^{2 \to 1}}{dyd^2p_T} \not= 0$ and
$\frac {dN_{\rm the}^{2 \to 2}}{dyd^2p_T} \not= 0$ indicates that
charmonia yielded in the prethermal stage and in the thermal stage overcome
the loss of charmonia due to the dissociation of charmonia in collisions with
gluons and hadrons. Now the question left is why the ratio $R_{AA}>1$ can 
take place at large $p_T$? 

Momentum and space distributions of partons 
in the prethermal stage were studied in detail in Refs. \cite{EW,LMW,LG}. 
The Fig. 5 given by Eskola and Wang \cite{EW} showed the variation of
transverse momentum distribution  $dN/d^2p_T$ with time. Before hard
scatterings partons are in Gaussian distribution due to the initial state
radiation. However, the large momentum transfer in the hard scatterings
considerably increases the parton numbers at large $p_T$ and an approximate
exponential
distribution comes with a larger $p_T$ tail. The abundance of partons with
transverse
momenta greater than 5 GeV is exactly a requisite what we want for getting the
enhancement of $J/\psi$ production at large $p_T$ as the $2 \to 1$ processes
$a+b \to c\bar c$ explicitly lead to large-$p_T$ $c\bar c$ pairs.
The dashed, dot-dashed and dotted curves in Figs. 2-3 stand for
direct charmonia from the initial nuclear collision, the prethermal stage and
the thermal stage, respectively. We found that
the yield of charmonia resulting from $c\bar c$ pairs
produced in the prethermal stage can be larger than that in the initial nuclear
collision and can be much larger than that in the thermal stage. 
Therefore, quark-gluon matter in the perthermal stage dominates the 
contributions to $R>1$.

We have seen that $R>1$ in the region $2.5~{\rm GeV}<p_T<7~{\rm GeV}$ at $y=0$
is a result of $c\bar c$ yielded from the prethermal stage and by means of the 
recombination mechanism. 
In Cu-Cu collisions at $\sqrt {s_{NN}}=200$ GeV the
thermal stage is shortened or disappeared 
and the number density of deconfined matter gets smaller. 
Hence, charmonium dissociation gets weaker and less $c\bar c$ pairs are
produced. But the two factors compete. Since quarks and gluons 
at high $p_T$ in the prethermal stage are still abundant, we can expect
$R_{AA} \sim 0.9$ or even larger than 1 as a result of deconfined matter in 
the prethermal stage as well as the recombination mechanism.

In summary, the enhancement of $J/\psi$ at high $p_T$ at midrapidity is related
to the earliest form of deconfined matter, i.e., quark-gluon matter in the 
prethermal stage. The nuclear modification factor $R_{AA} \sim 0.9$ or even
larger is due to $c\bar c$ produced in deconfined matter in the prethermal
stage and the recombination of $c$ and $\bar c$.

\vspace{0.5cm}
\leftline{\bf Acknowledgements}
\vspace{0.5cm}
This work was supported by National Natural Science Foundation
of China under Grant No. 10675079.

\noindent
\begin{figure}[htbp]
\centering
\includegraphics[scale=0.7,angle=270]{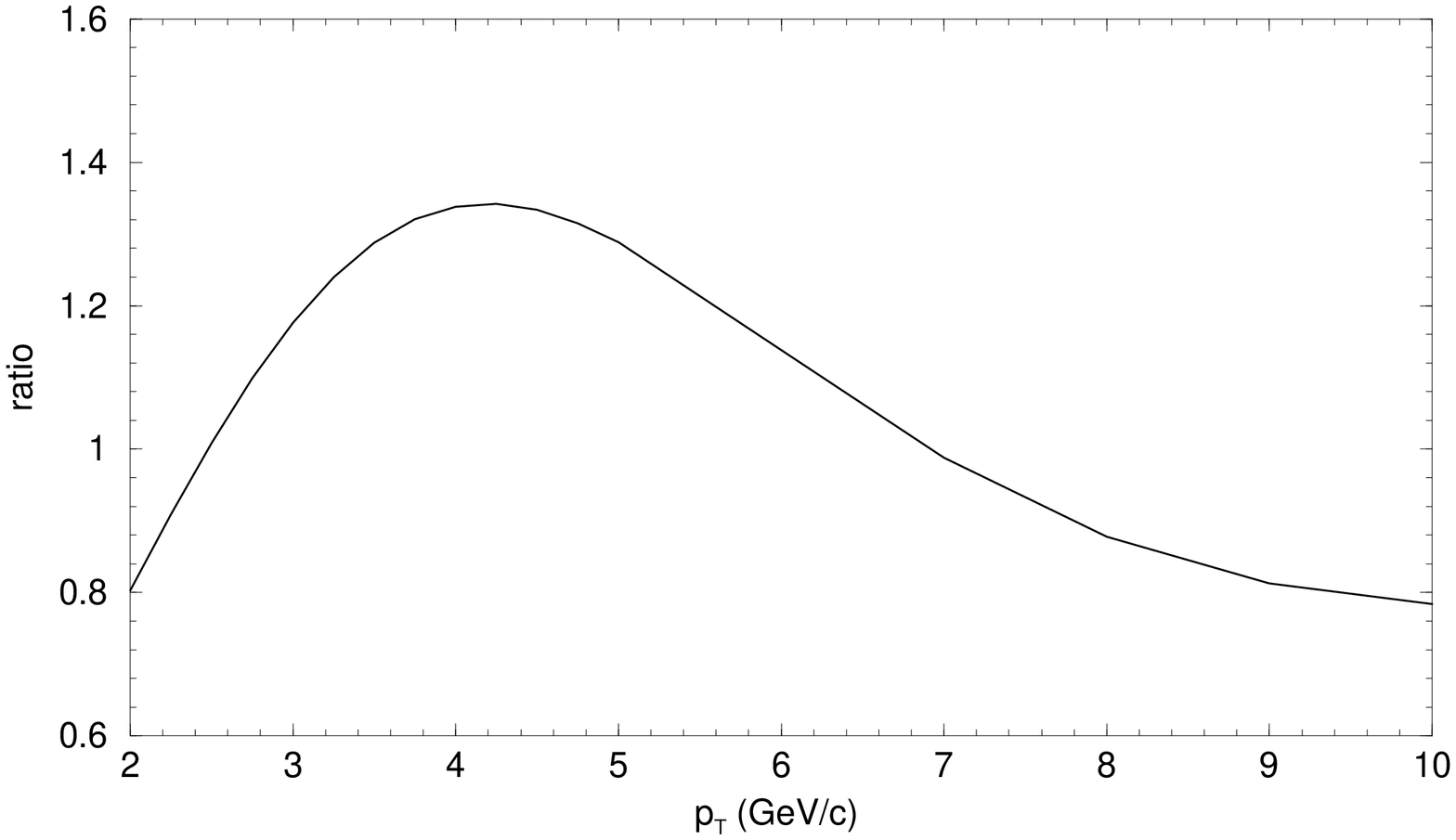}%
\caption{Ratio versus transverse momentum at rapidity $y=0$ 
for prompt $J/\psi$ production in central Au-Au collisions at 
$\sqrt {s_{NN}}=200$ GeV.}
\label{fig1}
\end{figure}

\begin{figure}[htbp]
\centering
\includegraphics[scale=0.8]{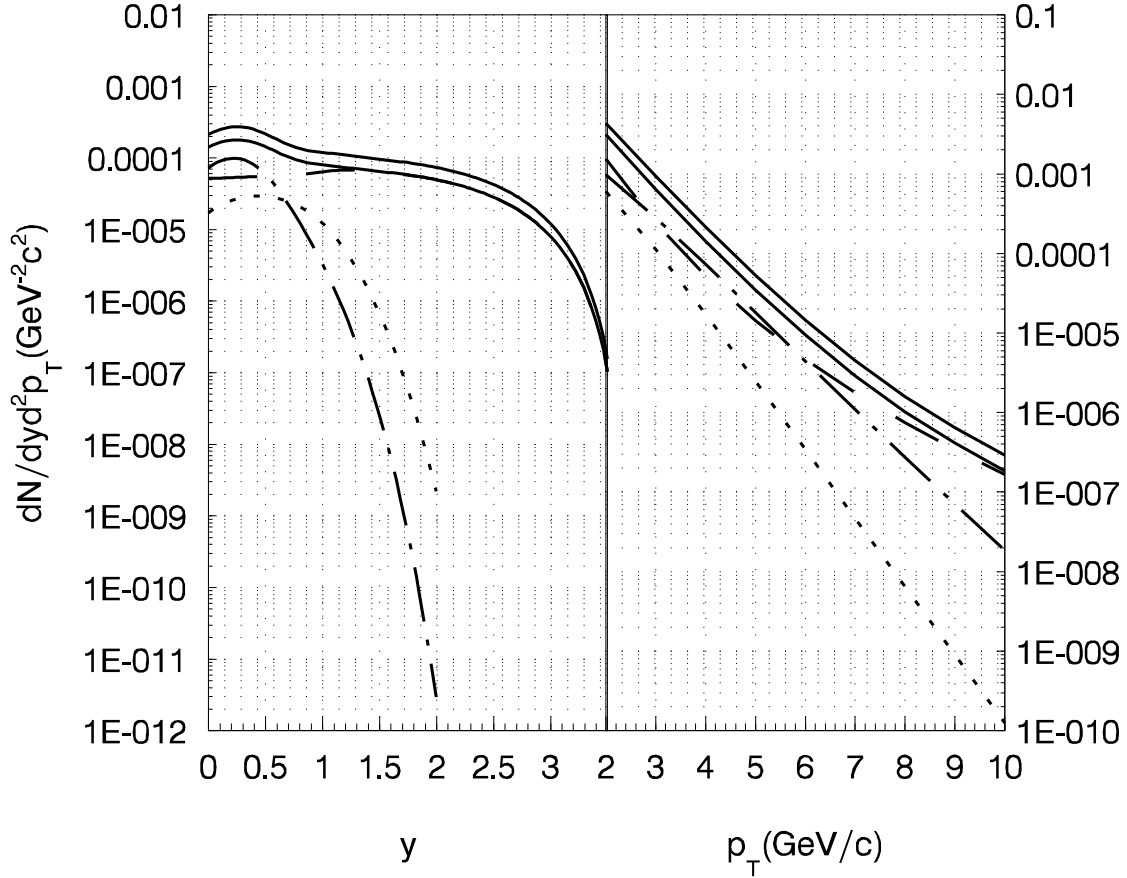}
\caption{$J/\psi$ momentum distributions versus rapidity at $p_T=4$ GeV in the 
left panel and transverse momentum at $y=0$ in the right panel for central 
Au-Au collisions at $\sqrt {s_{NN}}=200$ GeV. The dashed, dot-dashed, dotted 
and lower solid curves correspond to $c\bar c$ productions in the initial
collision, the prethermal stage, the thermal stage and the all three stages
(direct $J/\psi$), respectively. The upper solid curves (prompt $J/\psi$)
are the sum of all contributions including the
radiative feeddown from direct $\chi_{cJ}$ and the decay of direct $\psi'$.}
\label{fig2}
\end{figure}

\begin{figure}[htbp]
\centering
\includegraphics[scale=0.8]{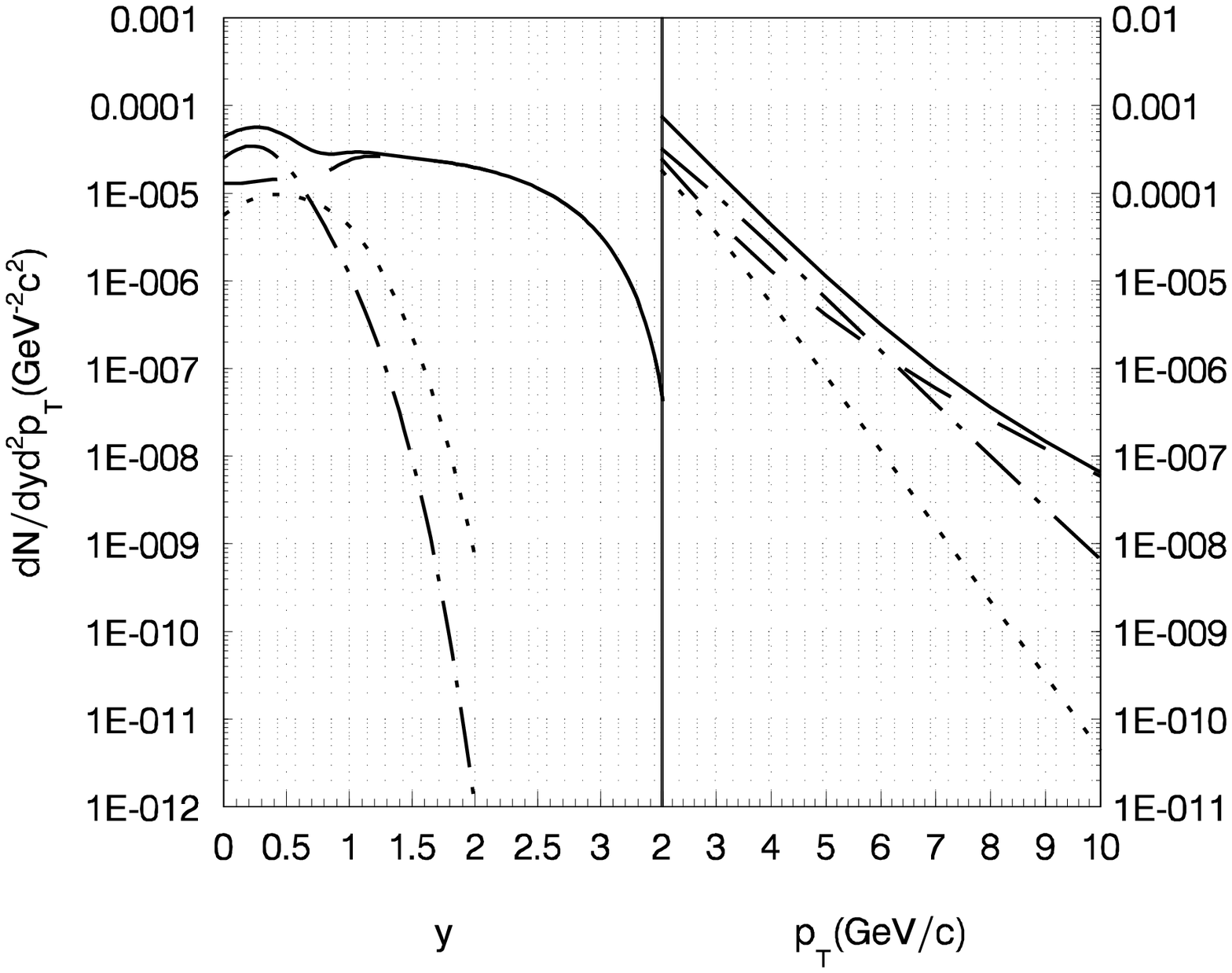}%
\caption{Direct $\psi'$ momentum distributions versus rapidity at $p_T=4$ GeV 
in the left panel and transverse momentum at $y=0$ in the right panel for 
central Au-Au collisions at $\sqrt {s_{NN}}=200$ GeV. 
The dashed, dot-dashed, dotted and solid
curves correspond to $c\bar c$ productions in the initial
collision, the prethermal stage, the thermal stage and the all three stages
(direct $\psi'$), respectively.}
\label{fig3}
\end{figure}

\begin{figure}[htbp]
\centering
\includegraphics[scale=0.8]{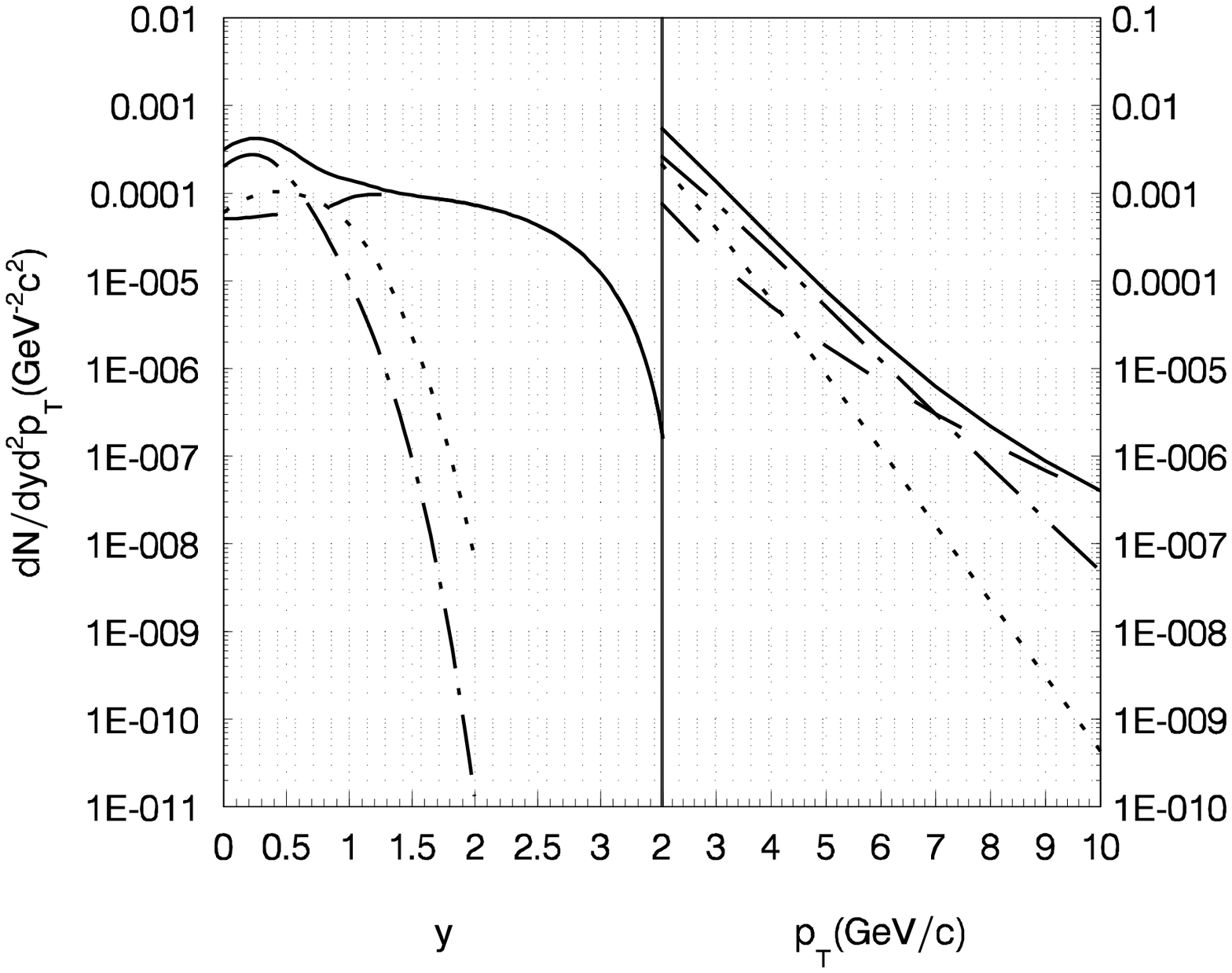}
\caption{Direct $\chi_{c}$ momentum distributions versus rapidity at $p_T=4$
GeV in the left panel and transverse momentum at $y=0$ in the right panel for 
central Au-Au collisions at $\sqrt {s_{NN}}=200$ GeV. 
The dashed, dot-dashed, dotted and  solid
curves correspond to $c\bar c$ productions in the initial
collision, the prethermal stage, the thermal stage and the all three stages
(direct $\chi_{c}$), respectively.}
\label{fig4}
\end{figure}

\end{document}